\documentclass[manuscript]{aastex}
\newcommand{\be}{\begin{equation}}
\newcommand{\ee}{\end{equation}}
\newcommand{\bse}{\begin{subequations}}
\newcommand{\ese}{\end{subequations}}
\newcommand{\bary}{\begin{eqnarray}}
\newcommand{\eary}{\end{eqnarray}}

\shorttitle{Long and short high energy emission in GRB090926A}
\shortauthors{Sacahui et al.}

\begin{document}
\title{The long and the short of the high energy emission in GRB090926A: an external shock }
\author{J. R. Sacahui \altaffilmark{1},  N. Fraija\altaffilmark{1}, M. M. Gonz\'alez\altaffilmark{1} and  W. H. Lee\altaffilmark{1}}
\affil{Instituto de Astronom\'ia, Universidad Nacional Aut\'{o}noma de M\'{e}xico, Apdo. Postal 70-264, Cd. Universitaria, M\'exico DF, 04510}
\email{ jsacahui@astro.unam.mx, nifraija@astro.unam.mx, magda@astro.unam.mx, wlee@astro.unam.mx}

\begin{abstract}
SSC emission from a reverse shock has been suggested as the origin for the high energy component lasting 2~s in the prompt phase of GRB98080923 \citep{fra12}. The model describes spectral indices, fluxes and the duration of the high-energy component as well as a long keV tail present in the prompt phase of GRB980923. Here, we present an extension of this model to describe the high-energy emission of GRB090926A. We argue that the emission consist of two components, one with a duration less than 1s during the prompt phase, and a second, longer-lasting GeV phase  lasting hundred of seconds after the prompt phase. The short high-energy phase can be described as SSC emission from a reverse shock similar to that observed in GRB980923, while the longer component arises from the forward shock. The main assumption
is that the jet is magnetized and evolves in the thick-shell case, and the calculated fluxes and  break energies are all consistent with the observed values. A comparison between the resulting parameters obtained for GRB980923 and GRB090926A suggests differences in burst tails that could be attributable to the circumburst medium, and this could account for previous analyses reported in the literature for other bursts.  We find that the density of the surrounding medium inferred from the observed values associated to the forward shock  agrees with standard values for host galaxies such as the one associated to GRB090926A.

\end{abstract}


\section{Introduction}

High-energy  gamma-ray emission from IC and SSC processes in different locations of a relativsitic jet have been  widely explored as a source for the observed high-energy emission in gamma-ray bursts.  Considering electrons accelerated in external (as the jet interacts with the circumburst medium) and internal (within the jet as the Lorentz factor of the flow varies) shocks and  different photon populations,  IC emission has been  discussed in detail in GRB internal shocks \citep{pap96,pil98, pan00a},  forward shocks \citep{sar96,tot98,wax97,pan98,wei98,chi99,der00a,der00b,pan00b} and  reverse shocks \citep{wan01a,wan01b, pee06}.   On the other hand,  SSC processes  from forward  \citep{sar01,wan01a} and reverse shocks \citep{gra03,wan01a,wan01b} have been investigated  separately to explain some atypical high energy components.   Also, the high-energy gamma-ray emission has been  described  using afterglow synchrotron radiation with a sufficiently low ISM density \citep{liu11,he11}. 

Recently, \citet{fra12} proposed that  the smooth tail  and  the high-energy component present in GRB980923 can be explained  in a unified manner with a SSC reverse and forward shock model.  While the smooth tail comes from  forward shock synchrotron emission,  the high-energy component  arises from synchrotron self-Compton (SSC) from the reverse shock.  In order to unify both processes  \citet{fra12} considered  different equipartition parameters for the reverse and forward shocks ( $\epsilon_{B,r}\neq \epsilon_{B,f}$ and $\epsilon_{e,r}\neq \epsilon_{e,f}$  ), thus leading to the requirement that the ejecta has to be highly magnetized. This assumption has been considered before, as in the work by \citet{kp03} for GRB 021211. 

We note that in previous work, \citet{kb09,kb10} and \citet{gh10} have explored how synchrotron radiation from a forward shock could account for long lasting high energy emission, finding good evidence in support of this model for a number of GRBs detected by the Fermi high energy detector (LAT).  More recently,  \citet{ver12} investigated possible scenarios to  obtain more than one observable component detected in the LAT energy range.  They explored SSC processes from forward and reverse shocks as possible mechanisms to produce the high-energy emission, and concluded that depending on the equipartition parameters  ($\epsilon_{B,r}=\epsilon_{B,f}$ and $\epsilon_{e,r}=\epsilon_{e,f}$), one or the other could develop to explain the energy component detected by LAT, besides requiring a population of photons from the prompt emission.

Here we extend the model used for GRB980923, showing that only the assumption of a magnetized jet is required to explain the complete high-energy emission with an SSC reverse and forward shock model. We also discuss the implications of the results with respect to to the circumburst medium.

\section{GRB 090926A}

GRB 090926A was reported as  a bright and long gamma-ray burst, observed by Fermi LAT/GBM \citep{ack10} at 04:20:26.99 (UT) on 26 September 2009. The GBM triggered on and localized it at RA=354.5$^\circ$  and Dec=-64.2$^\circ$ in J2000 coordinates.   GRB090926A was localized at $\approx$52$^\circ$ with respect to the pointing/axis direction and well within the field of view of Fermi.  It was independently detected by INTEGRAL SPI-ACS \citep{bi09}, Suzaku/WAM \citep{nod09}, CORONAS-PHOTON \citep{cha09} and the Konus-wind experiment \citep{gol09}.  The lightcurve observed by Konus-wind shows a prompt phase of 16~s followed by a weak tail up to  $\sim T_0+50$ s, and Suzaku WAM and CORONAS-PHOTON report a tail of similar energy range and duration.

Based upon the GCN report of the LAT detection, an  observation by the Swfit XRT and UVOT was performed  and an afterglow was detected at T$_0$+47 ks. Skynet/PROMPT also detected an optical afterglow emission 20~hr post burst \citep{nod09}. VLT observations using the X-shooter spectrograph determined a redshift of z=2.1062 \citep{mal09} and 
%the VLT/X-shooter spectroscopy of the GRB090926A afterglow 
concluded that the host galaxy associated with this burst was a dwarf irregular galaxy with one of the lowest metallicities observed for a GRB host galaxy \citep{del10}, $[X/H]=3.2 \times 10^{-3}-1.2 \times 10^{-2}$ with respect to solar.

Joint LAT and GBM data analysis concluded that GRB090926A presents a distinct  high energy power law component separated from the known BAND function \citep{ack10}.  The burst is fit  by a Band+CUTPL function with a high energy spectral break at E$_f$=$-1.41^{+0.22}_{-0.42}$ GeV and with a power law index of $-1.72\pm^{+0.10}_{-0.02}$. The fit was made for several time episodes noting that the extra power law component is very significant in one short episode lasting 0.15~s and in the subsequent episodes. The short episode is in coincidence with a sharp spike apparent  in the light curve above 100~MeV at $\sim t_{0}+10$~s. \citet{ack10} described  this short  peak through a power law with a cutoff energy  and obtained a value for the cutoff energy of E$_f$ = $0.40^{0.13}_{-0.06}$~stat.~$\pm 0.05$~syst.~GeV. The flux between 10~keV and 10~GeV for this short episode is $22.29\pm1.60 \times10^{-6}$~erg~cm$^{-2}$~s$^{-1}$ and the power law index is $\lambda = -1.71^{+0.02}_{-0.05}$. After the short spike the burst is better described by a Band+PL function with a power law index of $\lambda = -1.79\pm 0.03$ and a flux of $5.83\pm0.30 \times10^{-6}$~erg. Using the results of the Band+CUTPL model, \citet{ack10} infered an isotropic energy for this burst of E$_{iso}$ = $2.24\pm0.04\times10^{54}$~erg  in agreement with the value reported by \citet{gol09}. 

Given the similarities, and relevant differences between the two bursts cited, we summarize in  Table~1  the relevant observational parameters for both GRB090926A and GRB980923.  The presence of a short peak with high energy emission within the prompt phase in both events is remarkable.  We thus investigate the possibility of explaining the long duration emission at the highest energies with an extension of the model previously used for GRB980923 \citep{fra12}. 

\begin{center}
\begin{tabular}{ l c c }
  \hline \hline
 \scriptsize{Parameter} & \scriptsize{GRB980923} & \scriptsize{GRB090926A} \\
 \hline
 \scriptsize{$T_{90}$} & \scriptsize{32 s} & \scriptsize{20 s}\\
 \scriptsize{Duration of the High energy component} & \scriptsize{$<$ 2s} & \scriptsize{$<$ 1 s} \\
 \scriptsize{Spectral index of the tail (keV)}  & \scriptsize{2.4}  &  \scriptsize{ - }\\
 \scriptsize{Spectral index of the short high component}  & \scriptsize{$- 1.44 \pm 0.07$ \citet{gon11}}  &  \scriptsize{$-1.71^{+0.02}_{-0.05}$ }\\
 \scriptsize{Isotropic Energy  (erg)}  & \scriptsize{-}  &  \scriptsize{$2.2\times 10^{54}$}\\
 \scriptsize{Redshift }  & \scriptsize{-}  &  \scriptsize{$2.1096$ }\\
\scriptsize{Energy range of the tail}                            & \scriptsize{$ \sim$ 100 keV }  &  \scriptsize {$ \sim$ 50 keV \citep{gol09}} \\
\scriptsize{Range of the short high energy component}    & \scriptsize{ $>$ 200 MeV }  &  \scriptsize{ $ >$ 400 MeV}\\

\scriptsize{Range of the long high energy component}    & \scriptsize{ - }  &  \scriptsize{ $ \sim$ 10 GeV}\\

\end{tabular}
\end{center}

\begin{center}
\scriptsize{\textbf{Table 1. Observed parameters for  GRB980923 and GRB090926A  }}\\
\scriptsize{}
\end{center}

\section{Dynamics of the External Shocks} 

\subsection{Long GeV Component from Forward Shock}

For the forward shock, we assume that electrons are accelerated in the shock to a power law distribution of Lorentz factors $\gamma_e$, with a minimum Lorentz factor $\gamma_m$: $N(\gamma_e)d\gamma_e\propto \gamma_e^{-p}d\gamma_e$, where $\gamma_e\geq\gamma_m$ and $\epsilon_{e,f}$ and $\epsilon_{B,f}$ are the  constant fractions  of the shock energy that is transfered  into the electrons and the magnetic field, respectively. Then
\begin{eqnarray}
\gamma_{m,f}&=&\epsilon_{e,f}\biggl(\frac {p-2} {p-1}\biggr) \frac {m_p} {m_e}\gamma_f\cr
&=&524.6 \,\epsilon_{e,f}\,\gamma_{\rm f}
\end{eqnarray}
where we assume a typical value of $p$ for a synchrotron forward emission of $p=2.4$, as observed in GRB980923 by \citet{gib99} and generally
assumed in \citet{ver12}.
Adopting the notation of \citet{sar98}  we compute the typical and cooling frequencies of the forward shock synchrotron emission  \citep{sar98} which are given by,
\bary\label{synforw}
E_{\rm m,f}&\sim& 10.13 \,  \biggl(\frac{1+z}{3}\biggr)^{-1}\,\epsilon_{e,f,-1}^2\,\epsilon^{1/2}_{B,f,-4.3}\,n^{1/2}_{f,1}\,\gamma^{4}_{f,600}\,{\rm  keV}\cr
E_{\rm c,f}&\sim& 141.7\, \biggl(\frac{1+z}{3}\biggr)^{-1}\,\biggr(\frac{1+x_f }{11}\biggr)^{-2}\,\epsilon^{-3/2}_{B,f,-4.3}\,n^{-5/6}_{f,1}\,E^{-2/3}_{54}\,\gamma^{4/3}_{f,600}\, {\rm  eV}\cr
F_{\rm max,f}&\sim& 3.8 \times 10^{-2}\, \biggl(\frac{1+z}{3}\biggr)\,\epsilon^{1/2}_{B,f,-4.3}\,n^{1/2}_{f,1}\,D^{-2}_{28.3}\,E_{54}\,{\rm \, Jy}\cr
t_{\rm tr,f}&\sim& 467.5\, \biggl(\frac{1+z}{3}\biggr)\,\biggr(\frac{1+x_f }{11}\biggr)^{2} \,\epsilon_{e,f,-1}^2  \,\epsilon^{2}_{B,f,-4.3}\,n_{f,1}\,E_{54}\,{\rm s}
\eary

\noindent where the convention $Q_x=Q/10^x$ has been adopted in cgs units throughout this document unless something else is specified.  $t_{\rm tr,f}$ is the transition time when the spectrum changes from fast cooling to slow cooling,  $D$ is the luminosity distance, $n_{f}$ is the ISM density, $E$ is the energy, the term $(1+x_f)$ was introduced because a once-scattered synchrotron photon generally has energy larger than the electron mass in the rest frame of the second-scattering electrons.  Multiple scattering of synchrotron photons can be ignored. $x_f$ is given by \citet{sar01} as:

\begin{equation}
x_f = \left\{ \begin{array} {ll} 
\frac{\eta \epsilon_{e,f}}{\epsilon_{B,f}}, & \mathrm{if \quad}
\frac{\eta \epsilon_{e,f}}{\epsilon_{B,f}} \ll 1, \\ 
\left(\frac{\eta \epsilon_{e,f}}{\epsilon_{B,f}}\right)^{1/2}, & \mathrm{if \quad}
\frac{\eta \epsilon_{e,f}}{\epsilon_{B,f}} \gg 1. 
\end{array} \right.
\end{equation}

\noindent where $\eta=(\gamma_{\rm c,f}/\gamma_{\rm m,f})^{2-p}$ for slow cooling and $\eta=1$ for fast cooling.\\
From eq. \ref{synforw}, we observe directly that $\nu_{\rm m,f}  \gg \nu_{\rm c,f}$, the break energy $E_{\rm m,f}\sim 10.13\,$keV and  $t_{\rm tr,f}\sim 467.5$s  
implying that the synchrotron emission could be observed in the the fast cooling regime within a hundred of seconds from the burst. Thus, if $ E_{peak} (t) = E_0 (t-t_0)^{-\delta}$, then $\delta \sim 1.5$ \citep{bin07}

\bary\label{sscf}
E^{(IC)}_{\rm m,f}&\sim& 9.15\, \biggl(\frac{1+z}{3}\biggr)^{5/4}\,\epsilon_{e,r,-1}^{4}\,\epsilon_{B,f,-4.3}^{1/2}\,\gamma^{4}_{f,600}\,n^{-1/4}_{f,1}\,E^{3/4}_{54}\,t_{f,2}^{-9/4}\, {\rm \ GeV}\cr
E^{(IC)}_{\rm c,f}&\sim& 18.4 \, \biggl(\frac{1+z}{3}\biggr)^{-3/4}\,\biggr(\frac{1+x_f }{11}\biggr)^{-4}\,\epsilon_{B,f,-4.3}^{-7/2}\,n^{-9/4}_{f,1}\,E^{-5/4}_{54}\,t_{f,2}^{-1/4}\, {\rm \ eV}\cr
F^{(IC)}_{\rm max,f}&\sim& 1.59 \times 10^{-4} \,\biggl(\frac{1+z}{3}\biggr)^{3/4}\,\epsilon_{B,f,-4.3}^{1/2}\,n^{1/4}_{f,1}\,D^{-2}_{28.3}\,E^{5/4}_{54}\,t_{f,2}^{1/4}\,{\rm \,Jy}
\eary

From eq. \ref{sscf} we observe that the break energies and $(\nu F)_{\rm max}=10.9\times 10^{-7}$~erg~cm${^2}$~s$^{-1}$ are consistent with the values given by  \citet{ack10}.

\subsection{ Hard component  from thick shell  Reverse Shock}

Following the formalism of \citet{fra12} with the appropiated  values for GRB090926A,  we obtain for the synchrotron contribution,

\bary\label{synrev}
E_{\rm m,r}&\sim& 165.04\, \biggl(\frac{1+z}{3}\biggr)^{-1}\,\biggl(\frac{\epsilon_{e,r}}{0.65}\biggr)^{2}\,\biggl(\frac{\epsilon_{B,r}}{0.125}\biggr)^{1/2}\,\gamma^{2}_{r,3}\,n^{1/2}_{r,1}\, {\rm \ eV}\cr
E_{\rm c,r}&\sim& 1.1\times 10^{-3}\,\biggl(\frac{1+z}{3}\biggr)^{3/2}\,\biggl(\frac{1+x+x^2}{6}\biggr)^{-2}\,\biggl(\frac{\epsilon_{B,r}}{0.125}\biggr)^{-7/2}\,n^{-3}_{r,1}\,E^{-1/2}_{54}\,\gamma^{-6}_{r,3}\,\biggl(\frac{T_{90}}{20s}\biggr)^{5/2}\, {\rm \ eV}\cr
F_{\rm max,r}&\sim& 3.4 \times 10^2 \biggl(\frac{1+z}{3}\biggr)^{7/4}\,\biggl(\frac{\epsilon_{B,r}}{0.125}\biggr)^{1/2} \,n^{1/4}_{r,1}\,D^{-2}_{28.3}\,E^{5/4}_{54}\,\gamma^{-1}_{r,3}\,\biggl(\frac{T_{90}}{20s}\biggr)^{-3/4}\,{\rm \,Jy}
\eary

and for the SSC contribution,

\bary\label{ssc}
E^{(IC)}_{\rm m,r}&\sim& 414.3\,  \biggl(\frac{1+z}{3}\biggr)^{-7/4}\,\biggl(\frac{\epsilon_{e,r}}{0.65}\biggr)^{4}\,\biggl(\frac{\epsilon_{B,r}}{0.125}\biggr)^{1/2}\,\gamma^{4}_{r,3}\,n^{3/4}_{r,1}\,E^{-1/4}_{54}\,\biggl(\frac{T_{90}}{20s}\biggr)^{3/4}\, {\rm \ MeV}\cr
E^{(IC)}_{\rm c,r}&\sim& 0.7\times 10^{-5}\, \biggl(\frac{1+z}{3}\biggr)^{3/2}\,\biggl(\frac{1+x+x^2}{6}\biggr)^{-4}\,\biggl(\frac{\epsilon_{B,r}}{0.125}\biggr)^{-7/2}\,n^{-3}_{r,1}\,E^{-1/2}_{54}\,\gamma^{-6}_{r,3}\,\biggl(\frac{T_{90}}{20s}\biggr)^{-5/2}\, {\rm \ eV}\cr
F^{(IC)}_{\rm max,r}&\sim& 2.6 \times 10^{-1} \biggl(\frac{1+z}{3}\biggr)^{9/4}\,\biggl(\frac{\epsilon_{B,r}}{0.125}\biggr)^{1/2}\,n^{3/4}_{r,1}\,D^{-2}_{28.3}\,E^{7/4}_{54}\,\gamma^{-2}_{r,3}\,\biggl(\frac{T_{90}}{20s}\biggr)^{-5/4}\,{\rm \,Jy}.
\eary

From eq. \ref{ssc} we observe that the break energies and $(\nu F)_{\rm max}=8.2\times 10^{-6}$~erg~cm$^{-2}$~s$^{-1}$ are again consistent with the values given by \citet{ack10} and quoted above in \S~3.1. They also are similar to the ones obtained for GRB980923 as well as the values of  $\epsilon_{f, e,B}$ and densities $n_{f}$.  In fact, the ability to describe the short high energy emission from the reverse shocks  depends mainly on the value of the equipartition parameters.  We can interpret this as a requirement of a highly magnetized jet, and deal with the density in what follows. 

\section{Discussion and conclusions}

We have presented a model that describes both the short and long high energy components of GRB090926A. The parameters used to  describe the short high-energy component are similar to the ones found for GRB980923, as expected because of the similarities in the emission parameters (mainly duration, lightcurve and spectral index) for both bursts. For this component, our model is very dependent on the  values of $\epsilon_{B,r}=0.125$ and $\epsilon_{e,r}=0.65$, and implies the presence of a highly magnetized jet. We note that numerical particle in cell simulations carried out by \citet{ss11} appear to limit the magnetization parameter to $\sigma \simeq \epsilon_{B}/10 \leq 10^{-3}$.  However, these simulations have not been run for a long enough time to rule out if electrons can be accelerated for high magnetization shocks. Moreover, studies of other GRB events \citep{mes11, bar11,  gh10} with external shocks have found similar values to those reported here to be consistent with observations. 

The tail of the prompt emission of GRB980923 is described in our model as synchrotron radiation from the forward shock at energies of hundreds of keVs. This means that if  an SSC  component is present, it would appear at energies greater than tens of TeVs, which would not explain the long high-energy component observed in GRB090926A. However, it was found that SSC emission can account for the long high-energy component of GRB090926A with similar  values for the equipartition parameters as in GRB980923, but with a higher circumburst density. 

To complement the information intrinsic to the prompt high energy emission, we have roughly estimated the ambient density within the host using typical parameters for a dwarf galaxy ($L\sim 300$~pc) and the column densities reported by \citet{del10} and \citet{rau10}, obtaining $n=8.85$ and $11.93$~cm$^{-3}$, respectively.  These values are consistent with the value required by our model. 

The duration of the long high-energy component is estimated as the time for which  $E_{f,m}^{IC}$ decreases below 2~GeV, obtaining a value of $\approx 100$~s, in agreement with FERMI observations. Moreover, the keV-tail observed by \citet{gol09} can be related to the expected long (hundred of seconds) keV-component resulting from the forward synchrotron emission. One more difference in the tails of GRB090926A and GRB980923 besides the energy range, is their cooling regime, fast or slow respectively. Thus, if these two bursts are representatives of different surrounding media, as the value of $n$ indicates, there should be bursts with long tails in fast cooling regime at energies of a few keV and with long tails in slow cooling regime at energies of hundreds of keV. Coincidentally,  \citet{bin07} and  \citet{con02} have studied tails of bursts observed by SWIFT and BATSE, respectively. \citet{bin07} found single-peaked tails with cutoff energy evolving in time as  $E=E_0(\frac{t-t_0} {t_0})^{-\alpha}$ where $\alpha=1.2-1.4$, while \citet{con02} describes the temporal decay of the averaged tail of 400 bursts as a power law with index of $\alpha=0.6$. Both studies are in agreement with our result. Also, the afterglow reported 22 hours after the burst \citep{del10}  can be explained as the evolution in time of the synchrotron contribution.

The solution for the forward shock case given in this work is not unique. There are two more cases  \citep{ver12,liu11,he11}  with similar equipartition parameters but different density values. When $n=1$~cm$^{-3}$, the long high-energy component is described by forward SSC, but the duration of the keV-tail has to be longer than the transition time, which in this case is of the order of a day, much longer than the observed by \citet{gol09}. The second case is when $n=10^{-4}$ cm$^{-3}$, then the long high-energy component is described by forward synchrotron, thus no tail at keV is expected. We note that bursts with a short duration high energy emission as observed in GRB980923 and GRB090926A are candidates to present optical flashes \citep{sar99a,sar99b} as well as being detected by very-high gamma-ray observatories with wide-field of view as HAWC \citep{abeysekara12}. 

In summary, we have presented an SSC emission and forward shock model to explain the high energy emission observed in GRB090926A. The presence of tails in the keV regime, their duration, and time evolution are consistent with the reported observations. According to our model, both the short and hard high energy emission within the prompt phase can be interpreted as a signature of the magnetization of the jet,  while the energy regime of a long lasting (hundred of seconds) high energy emission would provide an indication of the circumburst medium through the density, with a possible correlation to the host galaxy type.

\acknowledgments

This work is partially supported by DGAPA-UNAM grant  IN105211 and CONACyT grants 105033 and 103520.

\end{document}